\documentclass[12pt]{iopart}
\usepackage{mathptmx}
\usepackage{graphicx}
\usepackage{subfigure}

\begin{document}

\title[]{Feynman's Clock for open quantum systems}

\author{David G. Tempel \footnote{tempel@physics.harvard.edu} and Al\'an Aspuru-Guzik\footnote{aspuru@chemistry.harvard.edu}}

\address{Department of Chemistry and Chemical Biology, Harvard University,
12 Oxford Street, 02138, Cambridge, MA}

\begin{abstract}
We show that Feynman's Clock construction, in which the time-evolution of a closed quantum system is encoded as a ground state problem, can be extended to open quantum systems. In our formalism, the ground states of an ensemble of non-Hermitian Feynman Clock Hamiltonians yield stochastic trajectories, which unravel the evolution of a Lindblad master equation. In this way, one can use Feynman's Clock not only to simulate the evolution of a quantum system, but also it's interaction with an environment such as a heat bath or measuring apparatus. A simple numerical example of a two-level atom undergoing spontaneous emission is presented and analyzed.\end{abstract}

\vspace{2pc}
\noindent{\it Keywords}: Mathematical Physics, Quantum Computation, Open Quantum Systems

\maketitle

\section{Introduction}

The notion that quantum computers can efficiently simulate the time-evolution of quantum systems was originally pioneered by Feynman~\cite{Feynman_82, Feynman_85, Feynman_86} and has since spurred a plethora of experimental and theoretical work in the field of quantum computation and simulation. Feynman and later Kitaev~\cite{Kitaev_02} envisioned a quantum simulator, in which the simulated system is entangled with a clock particle and the entire history of the simulation is encoded as the ground state of a Hamiltonian. This construction often referred to as ``Feynman's Clock" is particularly appealing, since it enables simulation of time-dependent quantum mechanics on a quantum computer using a time-independent setup~\cite{McClean_86, Aharonov_08, Mizel_04}. Feynman's Clock has also been an important tool in proving theorems, such as the equivalence of the adiabatic and gate models of quantum computation~\cite{Aharonov_08}.

Feynman's Clock in it's original formulation is restricted to isolated quantum systems evolving unitarily. However, many quantum systems of interest in chemistry and physics are not isolated, but undergo energy exchange and decoherence due to interaction with a thermal environment or measuring apparatus. Examples include energy transfer in photosynthetic and excitonic complexes~\cite{Mostame_2012}, condensed phase spectroscopy and cavity quantum electrodynamics to name a few. In all these systems, there still exists sufficient quantum coherence that one expects quantum mechanics to be important, but interactions with the environment are certainly not negligible. 

In the present manuscript, we will present a construction analogous to Feynman's Clock, but which is applicable to \textit{open} quantum systems. This construction is useful for a number of reasons. First, by mapping the open-system dynamics onto a ground state problem, it becomes time-independent and variational. This can be used to develop computational methods, which simulate open quantum systems on classical computers. This use of Feynman's Clock for closed quantum systems was presented in~\cite{McClean_86}. Second, it is a useful tool for proving theorems about quantum computation, when the dynamics are no longer assumed to be unitary~\cite{vest_09}. In this way, many of the proofs that use the Feynman Clock for unitary evolution might be extendable to open quantum systems. Third, as large-scale quantum computing devices are experimentally realized, it may be possible to build a Feynman Clock as a quantum simulator. In fact, this is what Feynman originally had imagined. For many realistic applications, a quantum simulator would need to be able to not only simulate a quantum system, but also its interaction with an environment.

The manuscript is organized as follows. In section 2, we review the Feynman Clock for unitary evolution and also the stochastic unraveling of the Lindblad master equation. Section 3 presents the formal theory behind Feynman's Clock for the Lindblad master equation. In section 4, the formal theory is demonstrated with a numerical study of a two-level atom undergoing spontaneous emission. Section 5 provides a conclusion by discussing experimental implementations and extensions of the theory to non-Markovian systems. We have set $\hbar = 1$ throughout, unless specified otherwise.

\section{Background}

Our goal in Section 3 will be to construct an ensemble of Feynman Clock Hamiltonians, which will have as their ground states the stochastic trajectories that unravel a Lindblad master equation. As a prelude, in this section we will separately review the Feynman Clock for unitary evolution and the Stochastic Schrodinger Equation (SSE) method of evolving the Lindblad equation.

\subsection{Feynman's Clock}

Feynman's original clock construction assumes an ideal quantum simulator described by a wave function evolving under the time-dependent Schr{\"odinger equation, $i\frac{\partial}{\partial t} |\psi(t) \rangle = \hat{H} |\psi(t) \rangle$, whose solution is $|\psi(t) \rangle = e^{-i \hat{H} t} |\psi(0) \rangle $, given an initial state $|\psi(0) \rangle$. The entire simulation is then encoded as a superposition in a ``history state" $|\eta \rangle = \sqrt{\frac{\delta t}{T+\delta t}} \sum_{t=0}^{T} |\psi(t) \rangle \otimes |t \rangle$, where $|t \rangle$ denotes the state of an auxiliary quantum system known as the ``clock," used to keep track of the evolution over the time of the simulation from $t=0$ to $t=T$. The clock can be any quantum degree of freedom, discrete or continuous, that is entangled with the Hilbert space of the system we wish to simulate. By performing a projective measurement of the clock at a specific time of interest t, the history state collapses to the wave function at that time, $|\psi(t) \rangle$.

In the construction of Kitaev~\cite{Kitaev_02}, $|\eta \rangle$ can be encoded as the ground state of the Hamiltonian,
\begin{eqnarray}
 \mathcal{H} &=& \sum_{t=0}^{T- \delta t}  -U \otimes |t+\delta t \rangle \langle t| - U^{\dag} \otimes |t\rangle \langle t+\delta t| + |t\rangle \langle t | \nonumber \\ &+& |t+\delta t \rangle \langle t+\delta t | +(1-|\psi(0) \rangle \langle \psi(0)|) \otimes |0\rangle \langle 0|,
\label{unitary_clock}
\end{eqnarray}
where $U = e^{-i \hat{H} \delta t}$ and $\delta t$ is the time-step which represents the distance between sites of the clock, assumed to be discrete. The first four terms in eq.~\ref{unitary_clock} ensure the history state encodes the correct time evolution, while the last term enforces the correct initial state. It can be readily verified that $ \mathcal{H} |\eta \rangle =0$ and because $ \mathcal{H}$ is positive semidefinite, $|\eta \rangle$ is the unique ground state of  $ \mathcal{H}$ with eigenvalue 0. In general, while the ground state encodes the history state, excited states of the Feynman Clock do not have an obvious physical interpretation.

\subsection{Stochastic Unraveling of the Lindblad Master Equation}~\label{stoch}

We wish to use a Feynman Clock construction to simulate an open quantum system, described by a density matrix evolving under the  widely used Markovian Lindblad master equation~\cite{lindblad},
\begin{equation}
\frac{d}{dt} \hat{\rho}_s = i [\hat{\rho}_s, \hat{H}_s] - \frac{1}{2} \sum_m (\hat{C}_m \hat{C}_m^{\dag} \hat{\rho}_s +\hat{\rho}_s \hat{C}_m \hat{C}_m^{\dag}) +  \sum_m \hat{C}_m \hat{\rho}_s \hat{C}_m^{\dag}.
\label{Lindblad}
\end{equation}
Here, $\hat{\rho}_s$ is the density matrix of the system, $\hat{H}_s$ is the system Hamiltonian, $\hat{C}_m$ and $\hat{C}_m^{\dag}$ describe interactions of the system with it's environment. Because eq.~\ref{Lindblad} describes the evolution of a mixed-state density matrix and not a pure-state wave function, a straightforward generalization of the Feynman Clock (eq.~\ref{unitary_clock}) is not possible. However, the Stochastic Schr\"odinger equation (SSE) procedure deals directly with wave functions and therefore serves as an ideal starting point for extending Feynman's Clock to open quantum systems~\cite{Molmer_93, Dalibard_92, Cohen_89}. 

The SSE  simulates a set of m individual realizations of the open quantum system, whose wave functions $\{|\psi^i(t)\rangle \}$ yield the ensemble averaged density matrix in eq.~\ref{Lindblad} by averaging over stochastic trajectories according to, \begin{equation} \hat{\rho}_s(t) = \frac{1}{m}\sum_{i=1}^{m} |\psi^i(t)\rangle \langle \psi^i(t) |. \label{avg} \end{equation} It can be shown that in the limit $m\rightarrow \infty$, eq.~\ref{avg} converges to the exact density matrix~\cite{Cohen_89}. Each realization can be thought of as simulating a single experiment on an individual member of the ensemble, such as a single-molecule fluorescence measurement or single-atom photon detection experiment.
The set of wave functions $\{|\psi^i(t)\rangle \}$ are simulated as follows. At time t, one evolves with a non-hermitian Hamiltonian to $t+\delta t$ according to  \begin{equation} |\psi^i(t+\delta t) \rangle = (1-i \hat{H}_s \delta t-\frac{1}{2} \sum_m \hat{C}_m^{\dag} \hat{C}_m) \frac{|\psi^i(t) \rangle}{\sqrt{1-\delta p(t)}},~\label{freed}\end{equation}with probability $1 - \delta p(t)$, where $\delta p(t) = \sum_m \delta p_m(t)$ and $\delta p_m(t) = \delta t \langle \psi^i (t)| \hat{C}_m^{\dag} \hat{C}_m | \psi^i(t) \rangle$. On the otherhand, with probability $\delta p(t)$, one instead collapses the wave function to the state
\begin{equation} 
|\psi^i(t+\delta t) \rangle = \frac{\hat{C}_m | \psi^i(t) \rangle}{\sqrt{ \delta p_m(t)/ \delta t}},
\label{jump}
\end{equation}
chosen from among the bath operators $\hat{C}_m$ with probability $\delta p_m(t)$. $\delta p(t)$ is typically small, so the majority of the time the system evolves freely without interaction with the environment. The Hamiltonian is non-Hermitian, because by learning that the system has not interacted with the environment, we have gained information, i.e. indirectly measured the system. Occasionally, with a small probability $\delta p(t)$, the system interacts with its environment causing wave function collapse.

\section{The Stochastic Feynman Clock - Formal Theory}

In this section we develop a formalism where history states are constructed, which correspond to the non-deterministic evolution of the stochastic trajectories discussed in section~\ref{stoch}. Each history state is encoded as the ground state of a non-Hermitian Hamiltonian, chosen probabilistically to enforce the correct stochastic jump probabilities. By ensemble averaging these history states, one can recover the entire history of the density matrix in eq.~\ref{Lindblad}. Before including stochastic jumps, however, we will begin with a deterministic, but non-Hermitian clock describing the free evolution in eq.~\ref{freed}.

\vspace{.5cm}

\subsection{The Non-Hermitian Feynman Clock}~\label{non-her-clock}

Our goal is to construct a history state, $|\eta \rangle  = \sqrt{\frac{\delta t}{T+\delta t}} \sum_{t=0}^{T} |\psi(t) \rangle \otimes |t \rangle$, where $|\psi(t) \rangle$ is identical to the evolution described by eq.~\ref{freed}.
Such a history state would encode the non-Hermitian evolution of an open system on the interval [0,T], when no stochastic jumps are generated. For an atom-photon experiment for example, $|\eta \rangle$ encodes the history of an atom, which is observed to evolve freely with no photons emitted. This history state is explicitly given by,
\begin{equation}
|\eta \rangle  = \sqrt{\frac{\delta t}{T+\delta t}} \sum_{t=0}^{T} \frac{(R)^{t}}{\sqrt{1-\delta p(t)}}|\psi(0) \rangle \otimes |t \rangle,
\label{non}
\end{equation}
where $R= 1 - i\hat{H}_s \delta t - \frac{1}{2} \delta t \sum_m \hat{C}_m^{\dag}  \hat{C}_m$, up to corrections of order $\delta t^2$. In order to construct a Feynman Clock, we need to write this history state as the ground state of a Hamiltonian. It can be readily verified that the history state in eq.~\ref{non} satisfies $\mathcal{H} |\eta \rangle =0$ where,
\begin{eqnarray}
\mathcal{H} &=& \sum_{t=0}^{T-\delta t} -R|t+\delta t \rangle \langle t| - R^{-1}|t\rangle \langle t+\delta t| + |t\rangle \langle t | \nonumber \\ &+& |t+\delta t \rangle \langle t+\delta t | +(1-|\psi(0) \rangle \langle \psi(0)|) \otimes |0\rangle \langle 0|
\label{non-clock}
\end{eqnarray}
and $R^{-1} = 1 + i\hat{H}_s \delta t + \frac{1}{2} \delta t \sum_m \hat{C}_m^{\dag}  \hat{C}_m$. The Hamiltonian in eq.~\ref{non-clock} is non-Hermitian, so it is not immediately obvious that $|\eta \rangle$ is the ground state or if this even has any meaning if the spectrum were complex. However, in the appendix we show that $\mathcal{H}$ has a complete, non-degenerate and real spectrum of positive eigenvalues. Therefore, $|\eta \rangle$ is in fact the ground state and we have succeeded in constructing a Feynman Clock, which encodes the free evolution of an open system when no jumps occur.

\subsection{The Stochastic Feynman Clock}~\label{exact}

So far we have constructed a non-Hermitian Feynman Clock describing the free SSE evolution, without stochastic jumps. We now show that it is possible to generate an ensemble of history states $\{ |\eta^i \rangle \}$ where,
\begin{equation}
|\eta^i \rangle= \sqrt{\frac{\delta t}{T+\delta t}} \sum_{t=0}^{T} |\psi^i(t) \rangle \otimes |t \rangle,
\end{equation}
is the history state of the ith stochastic trajectory and the set $\{ |\psi^i(t) \rangle\}$ are identical to those obtained from the SSE procedure in eq.~\ref{freed} and eq.~\ref{jump}. This is done by choosing an ensemble of Stochastic Feynman Clock Hamiltonians $\{ \mathcal{H}^{i} \}$ according to the following procedure. For each i, write $\mathcal{H}^{i}$ as a sum of local terms according to, 
\begin{equation}
\mathcal{H}^{i} = \sum_{t=0}^{T-\delta t} h^i(t+\delta t) + (1-|\psi(0) \rangle \langle \psi(0)|) \otimes |0\rangle \langle 0|.
\label{non-herm-jump}
\end{equation} 
Then for each value of t one chooses these terms to be,
\begin{eqnarray}
h^i(t+\delta t) &=&  -\sqrt{\frac{1-\delta p^i(t)}{1- \delta p^i(t+\delta t)}}R(\delta t)|t+\delta t \rangle \langle t| - \sqrt{\frac{1-\delta p^i(t+\delta t)}{1-\delta p^i(t)}}R^{-1}(\delta t) |t\rangle \langle t+\delta t| \nonumber \\ &+& |t\rangle \langle t | + |t+\delta t \rangle \langle t+\delta t |
\label{freest}
\end{eqnarray} 
with probability $1-p^i(t)$, where $\delta p^i(t) = \delta t \sum_m \langle \psi^i(t) | \hat{C}_m^{\dag}  \hat{C}_m | \psi^i(t) \rangle$. On the other hand, with probability $p^i(t)$ one instead chooses
\begin{equation}
h^i(t+\delta t) = (1 - \frac{\delta t}{\delta p^i_m} \hat{C}_m | \psi^i(t) \rangle \langle \psi^i(t) | \hat{C}_m^{\dag}) \otimes |t+\delta t \rangle \langle t+\delta t|
\label{jumpst}
\end{equation}
 from among the various bath operators $\hat{C}_m^{\dag}$ with probability $\delta p^i_m = \delta t  \langle \psi^i(t) | \hat{C}_m^{\dag}  \hat{C}_m | \psi^i(t) \rangle$. The terms in eq.~\ref{freest} force the ground state to have free evolution at the specified times, while the terms in eq.~\ref{jumpst} force the ground state to have the appropriate collapsed wave function corresponding to a jump. For a given Hamiltonian $\mathcal{H}^i$ generated in this way, it can be verified by direct substitution that the state $|\eta^i \rangle= \sqrt{\frac{\delta t}{T+\delta t}} \sum_{t=0}^{T} |\psi^i(t) \rangle \otimes |t \rangle$ satisfies $\mathcal{H}^i |\eta^i \rangle = 0$, provided the states $|\psi^i(t) \rangle$ are generated from the SSE with the same realization of jumps. The set of history states, $\{ |\eta^i \rangle \}$, encode the entire evolution of individual trajectories and can be used to compute ensemble averages. The density matrix of eq.~\ref{Lindblad} can be obtained by making projective measurements of the clock and then ensemble averaging these measurements according to,
 \begin{equation}
\hat{\rho}_s(t) =  \frac{1}{m} \sum_{i=1}^{m} Tr[ |\eta^i \rangle \langle \eta^i | |t \rangle \langle t| ] = \frac{1}{m}\sum_{i=1}^{m} |\psi^i(t)\rangle \langle \psi^i(t) |.
\label{density_matrix}
\end{equation}

 We see that the Hamiltonian in eq.~\ref{non-herm-jump} is a nonlinear functional of the state $|\eta^i \rangle$. This nonlinearity enters in two ways. First, there is an implicit nonlinearity, because the choice of terms in the Hamiltonian is determined probabilistically from the state $|\eta^i \rangle$. Second, there is an explicit dependence on  $|\eta^i \rangle$ appearing in the terms in eq.~\ref{freest} and eq.~\ref{jumpst}. In the appendix, we show that despite this nonlinearity,  the spectrum of each $\mathcal{H}^{i}$ is strictly real and there exists a corresponding $|\eta^i \rangle$ that is the ground state with eigenvalue 0. Because finding the spectrum of the Stochastic Feynman Clock is a nonlinear eigenvalue problem, it needs to be solved self-consistently to obtain an exact solution. However, in the next section we consider a perturbative expansion, which yields a valid approximation when the system-bath interaction is weak.

 \subsection{Perturbative Expansion of the Stochastic Feynman Clock}

As discussed in the previous section, the exact stochastic history states are solutions to a nonlinear eigenvalue problem, which must be solved self-consistently. However, it is often the case that the environment interacts only weakly with the system. In these situations, one expects that a majority of terms in the Hamiltonian will be of the form in eq.~\ref{freest} describing free evolution. Only occasionally does a jump occur, with a term of the form in eq.~\ref{jumpst} appearing. We can therefore linearize the Stochastic Feynman Clock and develop a perturbative expansion about the free evolution. 

One first solves the linear eigenvalue problem for the non-Hermitian Hamiltonian in eq.~\ref{non-clock}, and obtains the history state in eq.~\ref{non}, which describes free evolution when no jumps occur. We denote this state $ |\eta_0 \rangle$, which serves as the zeroth-order history state in our perturbative expansion. From $ |\eta_0 \rangle$, the set of jump probabilities $\{ \delta p_m(0) \}, \{ \delta p_m(\delta t) \}, \{ \delta p_m(2 \delta t) \}, ... \{ \delta p_m(T-\delta t ) \}$ for all bath operators m at each time-step are generated as follows. First, the initial set of jump probabilities $\{ \delta p_m(0) \}$ are obtained with a projective measurement of the clock particle at $t=0$ and simultaneous measurement of the bath operator $\hat{C}_m^{\dag}  \hat{C}_m$ according to $\delta p_m(0) = \delta t \langle \eta_0 | (\hat{C}_m^{\dag}  \hat{C}_m \otimes |0\rangle \langle 0|) |\eta_0 \rangle = \delta t \langle \psi(0)|  \hat{C}_m^{\dag}  \hat{C}_m | \psi(0) \rangle$. The remaining jump probabilities are obtained recursively, since the jump probabilities $\delta p_m(t + \delta t)$ at time $t+\delta t$, can be obtained from the history state $|\eta_0 \rangle$ and jump probabilities at earlier times $\{ \delta p_m(0) \}, \{ \delta p_m(\delta t) \}, ..., \{ \delta p_m(t) \}$ through the relation 
\begin{equation}
\delta p_m(t+\delta t) = \frac{\delta t}{1-\sum_{t'=0}^{t} \delta p(t')} \langle \eta_0| (\hat{C}_m^{\dag}  \hat{C}_m \otimes |t+ \delta t \rangle \langle t+ \delta t|) | \eta_0 \rangle, 
\label{recurs}
\end{equation}
where $t' < t$. This entire procedure necessitates storage of $M\frac{T}{\delta t}$ copies of the state $| \eta^0 \rangle$, where M is the number of bath operators, i.e. the algorithm is polynomial in the size of the system's Hilbert space and the run time of the simulation.

From the set of jump probabilities just obtained and ground state $|\eta^0 \rangle$, we now generate an ensemble $\{ \mathcal{H}^{i} \}$ of ``single-jump" Stochastic Clock Hamiltonians. This is done by writing the ith Hamiltonian in the ensemble as a sum of terms acting locally as in section~\ref{exact}, $\mathcal{H}^{i} \equiv \sum_{t=0}^{T-\delta t} h^i(t+ \delta t)$. Using the probabilities obtained in eq.~\ref{recurs}, we generate the ensemble by choosing,
\begin{eqnarray}
h^i(t+\delta t) &=&  -\sqrt{\frac{1-\delta p(t)}{1- \delta p(t+\delta t)}}R(\delta t)|t+\delta t \rangle \langle t| - \sqrt{\frac{1-\delta p(t+\delta t)}{1-\delta p(t)}}R^{-1}(\delta t) |t\rangle \langle t+\delta t| \nonumber \\ &+& |t\rangle \langle t | + |t+\delta t \rangle \langle t+\delta t |,
\label{freest2}
\end{eqnarray} 
with probability $\delta p(t) = 1-\sum_m \delta p_m(t)$. On the other hand, one chooses
\begin{equation}
h^i(t+\delta t) = (1 - \frac{\delta t}{\delta p_m(t)}) \hat{C}_m |\psi_0(t) \rangle \langle \psi_0(t)|\hat{C}_m^{\dag}   \otimes |t+\delta t \rangle \langle t+\delta t|
\label{jumpsaft}
\end{equation}
with probability $\delta p_m(t)$, where $|\psi_0(t) \rangle \equiv \langle t|\eta_0 \rangle$ . Once a jump has occurred, terms at later times are chosen to have the form in eq.~\ref{freest}, which enforces the one-jump approximation. The terms in eq.~\ref{freest2} force the ground state of $\mathcal{H}^{i}$ to have the correct free evolution before and after the jump, while the terms in eq.~\ref{jumpsaft} generate energy penalties that enforce the jumps with the correct probabilities. The Hamiltonian $\mathcal{H}^{i}$ generated this way is block diagonal, with each block corresponding to the free evolution before and after the jump has occurred. The ground state of $\mathcal{H}^{i}$ will have eigenvalue zero and be 2-fold degenerate, with one eigenstate corresponding to evolution before the jump and the other eigenstate to evolution after the jump. The physical history state, given as an equal superposition of these two degenerate states yields a single stochastic trajectory, $|\eta_1^i \rangle = \sqrt{\frac{\delta t}{T+\delta t}} \sum_{t=0}^{T} |\psi_1^i(t) \rangle \otimes |t \rangle $. The set of history states $\{ |\eta_1^i \rangle \}$ are first-order in the system-bath interaction. The set of wave functions $\{ |\psi_1^i(t) \rangle \}$ are precisely the wave functions from a SSE evolution, when only one jump has occurred. 

It is clear that the single-jump Stochastic Clock Hamiltonians depend only on $|\eta_0 \rangle$, so we need only solve a set of linear eigenvalue equations to to obtain the first-order history states $\{ |\eta_1^i \rangle \}$. Similarly, the set of first-order history states can be used to generate linear eigenvalue equations for a set of second-order history states $\{ |\eta_2^i \rangle \}$, which describe stochastic trajectories in which two jumps occur. The expansion can be continued up to order $N = \frac{T}{\delta t}$, which is equivalent to the solution of the nonlinear eigenvalue equations in section~\ref{exact}. In the appendix, we bound errors involved in truncating the expansion. Presently, we demonstrate the formal theory with an exactly solvable model system, in which the perturbation expansion to first-order is exact.

\section{Numerical Demonstration - A two-level atom undergoing spontaneous emission}

\begin{figure*}[htbp]
\begin{center}
 \includegraphics[width = 130mm, height = 100mm]{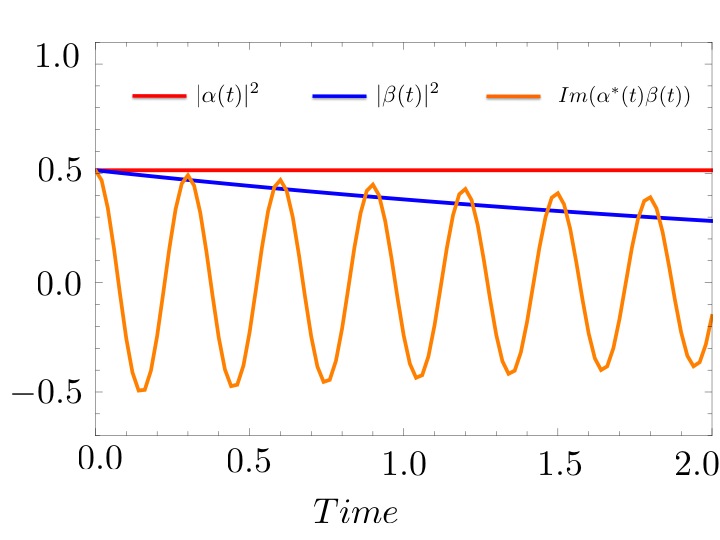}
\caption{\textbf{Ground state of the non-Hermitian clock} - Projections of the ground state $|\eta_0 \rangle$ of the deterministic non-Hermitian clock are plotted as a function of the ``time" parameter characterizing the clock states $|t \rangle$. $|\alpha(t)|^2 = |\langle t| \eta_0 \rangle|^2$ (red) is the ground state population of the atom and $|\beta(t)|^2 = |\langle t| \eta_0 \rangle|^2$ (blue) is the excited state population.}
\label{Figure1}
\end{center}
\end{figure*}

\begin{figure*}[htbp]
\begin{center}
 \includegraphics[width = 130mm, height = 100mm]{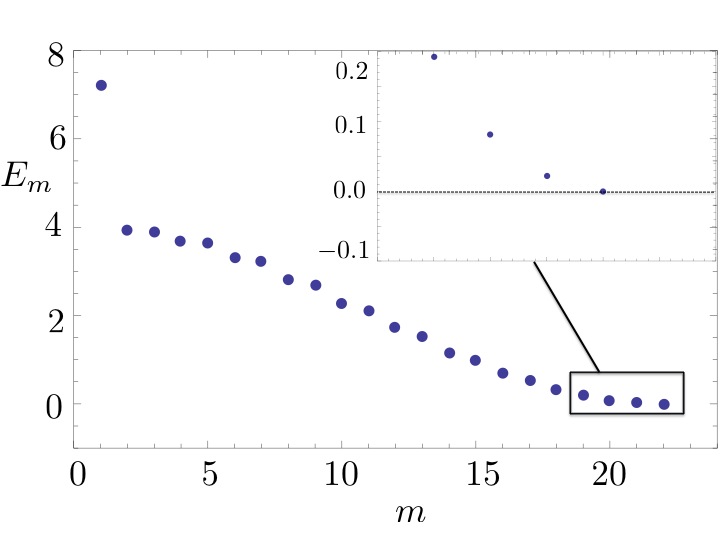}
\caption{\textbf{Spectrum of the non-Hermitian clock Hamiltonian} - The spectrum of the non-Hermitian clock Hamiltonian is non-degenerate, with a single zero eigenvalue corresponding to the ground state $|\eta_0\rangle$ (see inset). The spectrum is bounded between 0 and 4, except for a single state at energy 7.3 corresponding to an excited state that violates the initial condition imposed by the Hamiltonian, $(1-|\psi(0) \rangle \langle \psi(0)|) \otimes |0\rangle \langle 0|$.}
\label{Figure2}
\end{center}
\end{figure*}

\begin{figure*}[htbp]
\begin{center}
 \includegraphics[width = 130mm, height = 100mm]{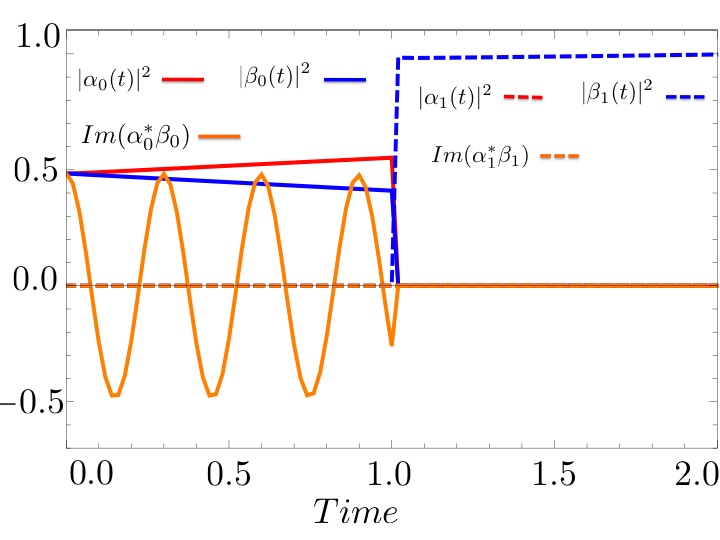}
\caption{\textbf{Ground state of a Stochastic Clock Hamiltonian with a single jump} - Populations and coherences of the two lowest eigenstates of a Stochastic Feynman Clock Hamiltonian, that implements a jump at $t=1.0 (\hbar \omega)$. The ground state is two-fold degenerate, with one of the eigenstates (solid lines) giving evolution before the jump and the other (dashed lines) giving evolution after.}
\label{Figure3}
\end{center}
\end{figure*}

As a simple numerical demonstration, we construct the Stochastic Feynman Clock for a two-level atom undergoing spontaneous emission, as determined by its interaction with a photon detector~\cite{Molmer_93, Dalibard_92}. If no photon is detected at time t, spontaneous emission has not occurred and the atom continues to evolve freely to time $t+ \delta t$. If on the other hand a photon is detected, spontaneous emission has occurred and the measurement causes the atomic wave function to jump to the ground state. In this situation the one-jump approximation is exact, because measurement causes the wave function to collapse to an eigenstate of the atomic Hamiltonian, leaving no possibility for a second jump to occur. The atomic wave function can be expanded in terms of the ground state $|0\rangle$ and excited state $|1 \rangle$ as $|\psi(t) \rangle = \alpha(t) |0\rangle + \beta(t) |1 \rangle$. The system Hamiltonian is then given by $\hat{H}_s = \omega |1\rangle \langle 1|$, where $\omega$ is the excitation energy. There is a single jump operator resulting from the measurement process given by $\hat{C} = \sqrt{\Gamma}|0\rangle \langle 0|$, where $\Gamma$ is the emission rate.

\begin{figure*}[htbp]
\begin{center}
 \includegraphics[width = 130mm, height = 100mm]{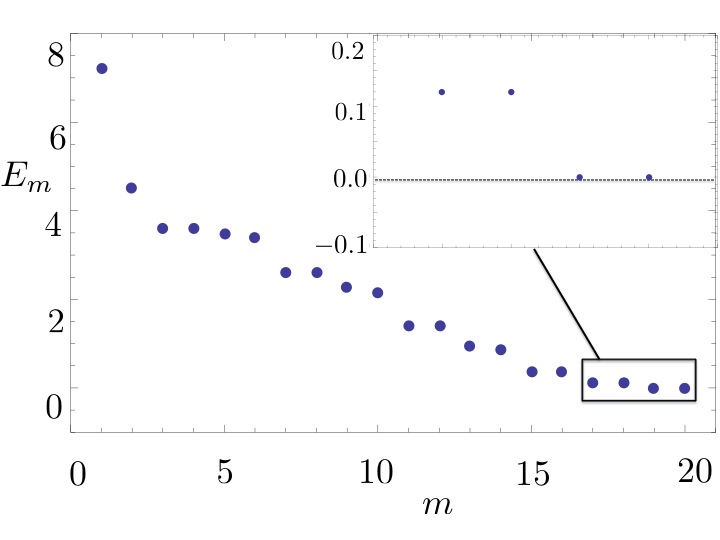}
\caption{\textbf{Spectrum of the Stochastic Clock Hamiltonian with a single jump} - With a single jump imposed, the spectrum splits into pairs of degenerate eigenstates, localized either before or after the jump. The ground state is two-fold degenerate, with eigenvalue 0.}
\label{Figure4}
\end{center}
\end{figure*}

\begin{figure*}[htbp]
\begin{centering}
$\begin{array}{c@{\hspace{1in}}c}
 \includegraphics[width = 130mm, height = 100mm]{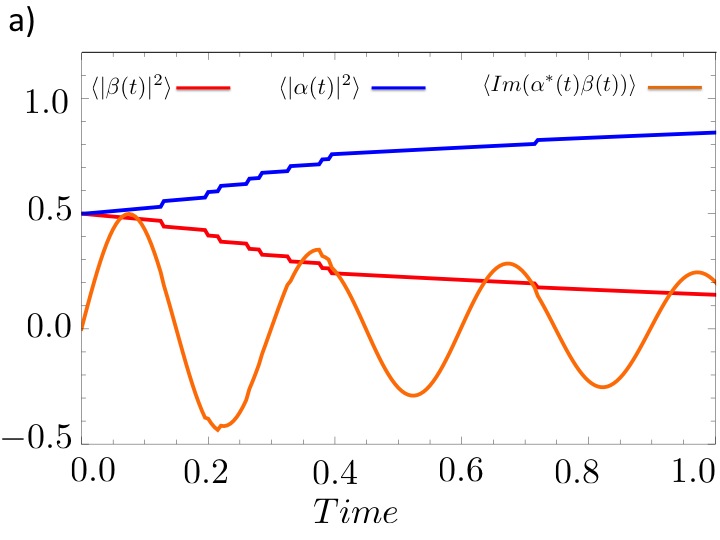}\\ 
 \includegraphics[width = 130mm, height = 100mm]{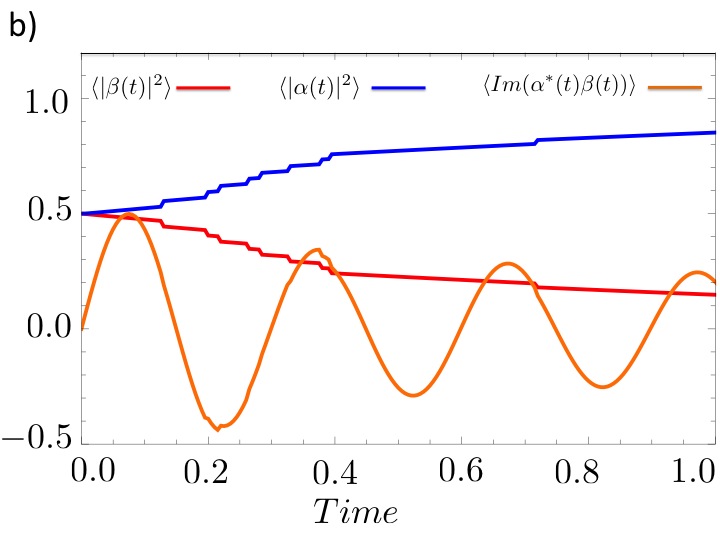}
 \end{array}$
 \end{centering}
\caption{\textbf{Density matrix of the Stochastic Feynman Clock and the Stochastic Schrodinger Equation} - a) The open system density matrix of a decaying two level atom obtained by ensemble averaging 20 history states generated by the Stochastic Feynman Clock procedure. b) The open system density matrix generated using the SSE. The two density matrices are nearly indistinguishable. In both cases the runtime of the evolution is $T= 1.0 ( \hbar \omega)$.}
\label{Figure5}
\end{figure*}

\begin{figure*}[htbp]
\begin{centering}
$\begin{array}{c@{\hspace{1in}}c}
 \includegraphics[width = 130mm, height = 100mm]{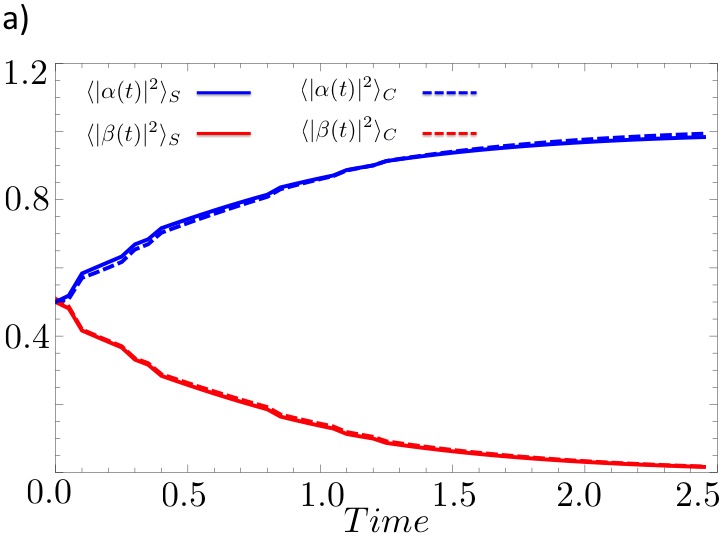}\\ 
 \includegraphics[width = 130mm, height = 100mm]{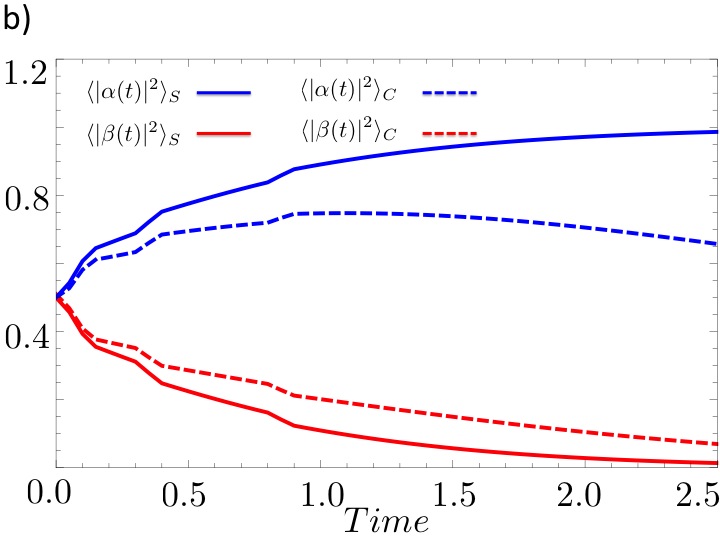}
 \end{array}$
 \end{centering}
\caption{\textbf{Open-system density matrix for different run times with static disorder} - a) Ground (blue) and excited state (red) populations of the density matrix obtained from the SSE (solid lines) and clock history state (dashed lines) for a runtime of $T= 2.5 (\hbar \omega)$. b) The same as in part a, but with a runtime of  $T= 10.0 (\hbar \omega)$}
\label{Figure6}
\end{figure*}

An ensemble of two-level atoms in this setup is described by the Lindblad master equation,
\begin{equation}
\frac{d}{dt} \hat{\rho}_s = i [\hat{\rho}_s, \hat{H}_s] - \frac{\Gamma}{2}(\hat{\sigma}^{+} \hat{\sigma}^- \hat{\rho}_s + \hat{\rho}_s \hat{\sigma}^{+} \hat{\sigma}^-) + \Gamma \hat{\sigma}^- \hat{\rho}_s \hat{\sigma}^+,
\label{min-Lind}
\end{equation}
where $\hat{\sigma}^+ = |1\rangle \langle0|$ and $\hat{\sigma}^- = |0\rangle \langle1|$. The above density matrix is obtained by averaging over stochastic trajectories of the SSE as described in section~\ref{stoch}.

For the simulation, we choose the initial state to be an equal superposition of the atomic ground state and first excited state, i.e. $\alpha(0) = \beta(0) = \frac{1}{\sqrt{2}}$. The deterministic non-Hermitian Clock Hamiltonian in eq.~\ref{non} is constructed and the ground state $|\eta_0 \rangle  = \sqrt{\frac{\delta t}{T+\delta t}} \sum_{t=0}^{T}  |\psi_0(t) \rangle \otimes |t \rangle$ is obtained via an exact numerical diagonalization of the Clock Hamiltonian, $\mathcal{H}_0$. The populations and coherence obtained from the coefficients $|\psi_0(t) \rangle = \langle t| \eta_0 \rangle$ are shown in Figure~\ref{Figure1} as a function of $|t\rangle$. As expected, the free non-Hermitian evolution causes the excited state population and coherence to decay, while the ground state population of the atom remains unchanged. Although we have solved for the ground state of a time-independent Hamiltonian, we get exactly what we would expect from a propagation of the SSE with no jumps.  Figure~\ref{Figure2} shows the spectrum of $\mathcal{H}_0$, which has a single non-degenerate zero eigenvalue corresponding to $|\eta_0 \rangle$. In general, the excited states of $\mathcal{H}_0$ are unphysical and only the ground state encodes a valid evolution, which is true of the unitary Feynman Clock as well~\cite{McClean_86}. Figure~\ref{Figure3} shows populations and coherences of a single Stochastic Feynman Clock, in which a jump penalty (eq.~\ref{jumpsaft}) corresponding to a photon detection has been imposed at $t=1.0 (\hbar \omega)$. The ground state is two-fold degenerate, with one ground state corresponding to free evolution before the jump and the other after the jump has caused the wave function to collapse to the atomic ground state. In Figure~\ref{Figure4}, one sees that in general, the spectrum is divided into pairs of degenerate states for every energy level. For each pair, one state is localized before the jump occurs, while the second is localized afterward. This happens because the clock Hamiltonian assumes a block diagonal form, with one block acting only on clock states $\{ |t\rangle \}$ before the jump and the other block acting only on states after the jump.

Figure~\ref{Figure5}a shows the elements of the density matrix obtained from averaging the ground states of 20 Stochastic Feynman Clock Hamiltonians, for a runtime of $T= 1.0 (\hbar \omega)$. The density matrix is seen to be nearly identical to that obtained from a conventional propagation of the SSE using the same realization of the random number generator, shown in Figure~\ref{Figure5}b. This serves to demonstrate that the Stochastic Feynman Clock exactly reproduces the unraveling of the Lindblad equation with the SSE (eq.~\ref{min-Lind}).

In a realistic experimental setup one expects imperfections in the Stochastic Clock Hamiltonian to cause the history state to deviate from the open-system dynamics simulated in real-time by a SSE. To study the effects of such imperfections, we include diagonal static disorder by choosing the Stochastic Clock Hamiltonians to have the form $\mathcal{H}^{i} + \mathcal{\delta}$, where $\delta$ is a positive-semidefinite diagonal random matrix. In Figure~\ref{Figure6}, we see that for small runtimes the static disorder has little effect and the density matrix produced by ensemble averaging Stochastic Clock history states still agrees faithfully with that from the SSE evolution. However, as the runtime increases, the density matrix from the Stochastic Clock history states are seen to deviate appreciably from the SSE evolution. This occurs due to contamination from excited states of the Stochastic Clocks, which do not correspond to physical evolution. Specifically, the gap between the ground and first excited history state of the clock is proportional to $\frac{1}{T^2}$~\cite{McClean_86}, so the effects of noise become appreciable when $\delta_{max}T^2 \sim 1$, where $\delta_{max}$ is the maximum eigenvalue of the matrix $\delta$. However, the error grows only polynomially in the runtime and in general static disorder can be minimized more easily than dynamic disorder. From a quantum simulation standpoint, this might offer an advantage of the Feynman Clock construction over conventional simulation in real time.

\section{Conclusion and Outlook}

Experimental implementation of the Stochastic Feynman Clock should be possible with a variety of highly tunable quantum systems, such as ultracold atoms~\cite{Bakr_09}, superconducting qubits~\cite{Hime_06} and quantum dots~\cite{Greve_11}. For instance, with ultracold atoms in an optical lattice, the optical lattice can be tuned to a very large local potential at a randomly chosen site to impose stochastic jumps. In contrast, a perfectly periodic optical lattice in the superfluid regime will have delocalized eigenstates corresponding to a history state with free evolution. Non-Hermitian evolution can be generated by coupling sites representing the history state to a large number of other sites in the lattice, effectively generating a continuum. Furthermore, many stochastic history states can be generated in parallel in a single experiment and stored in disconnected sites of the optical lattice. The coefficients of the history states can be read and manipulated as required by our procedure using a high resolution quantum gas microscope as presented in ref.~\cite{Bakr_09}.

From an experimental standpoint, the Feynman Clock formulation of quantum simulation offers a decided advantage over conventional simulation in real-time. By recasting the simulation as a time-independent problem, all quantum gates can be constructed as ground-state interaction terms and no ultrafast, real-time manipulations are needed. Of course, before experimentally implementing the Feynman Clock for open quantum systems, an experimental simulation of the unitary Feynman Clock will be a necessary prerequisite. 

Many theoretical results in quantum computation rely on the unitary Feynman Clock to prove theorems. We anticipate that the Stochastic Feynman Clock can be a useful tool for extending many of these results to open quantum systems.  Examples include the equivalence of the adiabatic and circuit models of quantum computing~\cite{Aharonov_08, Mostame_10} and the complexity of k-local Hamiltonians~\cite{Kempe_06}.
We have focused specifically on Markovian environments, but it is also possible to construct non-Markovian Stochastic Feynman Clocks starting from a non-Markovian quantum jump model~\cite{Rebentrost_09}. Also, we have chosen to work directly with wave functions by using the SSE as our starting point, rather than the master equation for the density matrix. This is the most straightforward generalization of the unitary Feynman clock, since one still works within a Hilbert space. However, this approach has the obvious disadvantage that the Stochastic Feynman Clock Hamiltonians are non-linear functionals acting in the Hilbert space. In future work, we will explore the possibility of formulating Feynman's Clock in Liouville space, which allows one to derive linear equations for the density matrix~\cite{Mukamel_95}.

\section{Acknowledgments}

We are grateful to Jarrod McClean, Joel Yuen-Zhou and Gian Giacomo Guerreschi for valuable discussions.  We acknowledge NSF CHE-1152291 for financial support.

\appendix

\section{Spectrum of the Non-Hermitian Clock}~\label{AppA}

In section~\ref{non-her-clock} we introduced the non-Hermitian clock $\mathcal{H}$ and found that the history state $|\eta \rangle  = \sqrt{\frac{\delta t}{T+\delta t}} \sum_{t=0}^{T} |\psi(t) \rangle \otimes |t \rangle$ satisfies $\mathcal{H} |\eta \rangle = 0$. Our goal is to show that the spectrum of $\mathcal{H}$ is real and positive so that $|\eta \rangle$ is in fact the ground state. 

We first show that the operator $\mathcal{H}$ is normal and therefore admits a spectral decomposition~\cite{Bachman_66}. We can write $\mathcal{H} = \mathcal{L} + \mathcal{T}$ as a sum of a Hermitian part $\mathcal{L}$  and anti-Hermitian part, $\mathcal{T}$. Explicitly,
\begin{eqnarray}
\mathcal{L}& =& \sum_{t=0}^{T-\delta t} -(1-i \delta t \hat{H})|t+\delta t \rangle \langle t| - (1+i \delta t \hat{H}) |t\rangle \langle t+\delta t| + |t\rangle \langle t | \nonumber \\ &+& |t+\delta t \rangle \langle t+\delta t| +(1-|\psi(0) \rangle \langle \psi(0)|) \otimes |0\rangle \langle 0|
\end{eqnarray}
and 
\begin{equation}
\mathcal{T} = \delta t \sum_{t=0}^{T-\delta t} \left[ \hat{D} |t+\delta t \rangle \langle t| - \hat{D} |t \rangle \langle t+\delta t | \right],
\end{equation}
where $ \hat{D}=\frac{1}{2} \delta t \sum_m \hat{C}_m^{\dag}  \hat{C}_m$.
Working out the commutator of $\mathcal{H}$ with its adjoint $\mathcal{H}^{\dag}$ and keeping only terms linear in $\delta t$ one finds,
\begin{equation}
[\mathcal{H}^{0},(\mathcal{H}^{0})^{\dag}] \approx \delta t \left[\hat{D}(1-|\psi_0 \rangle \langle \psi_0|)+ (1-|\psi_0 \rangle \langle \psi_0|) \hat{D} \right] \otimes |0\rangle \langle 0|.
\end{equation}
We notice 2 things. First, while the operator $\mathcal{H}$ is not normal strictly speaking, it becomes normal in the limit  $\delta t \rightarrow 0$. Second, the commutator is non-zero only if the bath operator $\hat{D}$ does not commute with the projector $P_{\psi0} \equiv |\psi_0 \rangle \langle \psi_0| \otimes I$. Consequently,
\begin{equation}
P_{\psi0} [\mathcal{H}^{0},(\mathcal{H}^{0})^{\dag}] P_{\psi0} = 0
\end{equation}
and $\mathcal{H}$ is normal in the subspace of history states satisfying the initial condition. Therefore, $\mathcal{H}$ has a complete spectrum of eigenstates $\{ |\eta_k \rangle \}$ satisfying,
\begin{equation}
\mathcal{H}^{0} |\eta_k \rangle = \lambda_k |\eta_k \rangle,
\label{eig}
\end{equation}
either in the limit $\delta t \rightarrow 0$ or to all orders in $\delta t$ within the projected space of $P_{\psi0}$.

We now prove that all of the eigenvalues $\lambda_k$ are real and positive. To this end, we introduce an operator $\hat{O}$, which acts on the clock states as $\hat{O} |t \rangle= (1+t \delta t  \hat{D}) |t \rangle$. It has an inverse whose action is $\hat{O}^{-1} |t \rangle= (1-t \delta t  \hat{D}) |t \rangle$. We may then re-write eq.~\ref{eig} as
\begin{equation}
 \hat{O} \mathcal{H}\hat{O}^{-1} \hat{O} |\eta_{k} \rangle = \lambda_k  \hat{O} |\eta_{k} \rangle,
 \label{proj2}
\end{equation}
where
\begin{eqnarray}
\hat{O} \mathcal{H}\hat{O}^{-1} &=& \sum_{t=0}^{T-} \left[ -(1 -i\hat{H}) |\delta t+\delta t \rangle \langle t| - (1 + i\hat{H} ) |t\rangle \langle t+\delta t| + |t\rangle \langle t | + |t+\delta t \rangle \langle t+\delta t | \right]  \nonumber \\ &+& (1-|\psi(0) \rangle \langle \psi(0)|) \otimes |0\rangle \langle 0|.
\end{eqnarray}
This transformed Hamiltonian is Hermitian, positive semidefinite and has eigenvalues between 0 and 4. Since it has an identical spectrum to the original Hamiltonian in eq~\ref{eig}, this is true of the Hamiltonian $\mathcal{H}$ as well.

\section{The Spectrum of the Stochastic Feynman Clock}~\label{AppB}

We now show that despite being nonlinear functionals of their respective ground states, the Stochastic Feynman Clock Hamiltonians $\{ \mathcal{H}^i \}$ have a real and positive spectrum. Consequently,  for each realization the state $|\eta^i \rangle= \sqrt{\frac{\delta t}{T+\delta t}} \sum_{t=0}^{T} |\psi^i(t) \rangle \otimes |t \rangle$, which satisfies $\mathcal{H}^i |\eta^i \rangle = 0$ is the ground state. To show this, consider a particular realization,
\begin{equation}
\mathcal{H}^{i} = \sum_{t=0}^{T-\delta t} h^i(t+\delta t) + (1-|\psi(0) \rangle \langle \psi(0)|) \otimes |0\rangle \langle 0|,
\label{non-herm-jump-app}
\end{equation} 
which has local jump Hamiltonians of the form,
 \begin{equation}
 h(s_j) = (1 - \frac{\delta t}{\delta p_m} \hat{C}_m | \psi(s_j- \delta t) \rangle \langle \psi(s_j - \delta t) | \hat{C}_m^{\dag})\otimes |s_j \rangle \langle s_j|
 \end{equation}
at a set of n times $\{s_j\}$ and free non-Hermitian evolution at other times. The Hamiltonian in eq.~\ref{non-herm-jump-app} can be written as,
\begin{equation}
\mathcal{H} = \sum_{j=0}^{n} \mathcal{H}(s_j),
\end{equation}
where
\begin{eqnarray}
\mathcal{H}(s_j) &=&  \sum_{t=s_j}^{s_{j+1}} \left[ -\sqrt{\frac{1-\delta p^i(t)}{1- \delta p^i(t+\delta t)}}R(\delta t)|t+\delta t \rangle \langle t| - \sqrt{\frac{1-\delta p^i(t+\delta t)}{1-\delta p^i(t)}}R^{-1}(\delta t) |t\rangle \langle t+\delta t| \right] \nonumber \\ &+& |t\rangle \langle t | + |t+\delta t \rangle \langle t+\delta t | +1 - \frac{\delta t}{\delta p_m} \hat{C}_m | \psi(s_j- \delta t) \rangle \langle \psi(s_j - \delta t) | \hat{C}_m^{\dag})\otimes |s_j \rangle \langle s_j|.
\end{eqnarray}
We see from the above expression, that each $\mathcal{H}(s_j)$ has the form of a free non-Hermitian clock Hamiltonian, starting from the initial state $\sqrt{\frac{\delta t}{\delta p_m}} \hat{C}_m | \psi(s_j- \delta t) \rangle$ instead of $|\psi(0) \rangle$. Furthermore, the various $\mathcal{H}(s_j)$ commute with one another. Therefore, $\mathcal{H}$ has a block diagonal structure, and the eigenvalue equation $\mathcal{H} |\eta_k \rangle = \lambda_k |\eta_k \rangle$ separates into separate eigenvalue equations in each block,
\begin{equation}
\mathcal{H}(s_j) |\eta_k(s_j) \rangle = \lambda_k(s_j) |\eta_k(s_j) \rangle.
\end{equation}
Applying the results of~\ref{AppA}, each $\lambda_k(s_j)$ is real and positive. Therefore, $\lambda_k = \sum_j \lambda_k(s_j)$ is real and positive as well. The state
 \begin{eqnarray}
 |\eta \rangle &=& \sqrt{\frac{\delta t}{T+\delta t}} \sum_{t=0}^{s_{0} - \delta t} \frac{\hat{R}^{\frac{t}{\delta t}}}{\sqrt{1-p(t)}}  |\psi(0) \rangle \otimes |t\rangle+ \sum_{j=0}^{n} \sum_{t=s_j}^{s_{j+1} - \delta t} \frac{\hat{R}^{\frac{t-s_j}{\delta t}}}{\sqrt{1-p(t)}} \hat{C}_m |\psi(s_j-\delta t) \rangle \otimes |t\rangle \nonumber \\ &+& \sum_{t=s_{n+1}}^{T} \frac{\hat{R}^{\frac{t-s_{n+1}}{\delta t}}}{\sqrt{1-p(t)}} \hat{C}_m |\psi(s_{n+1}- \delta t) \rangle \otimes |t\rangle ,
 \label{state}
 \end{eqnarray}
satisfies $\mathcal{H} |\eta \rangle = 0$, so it is the ground state. From inspection, we see that this state has exactly the form $|\eta^i \rangle= \sqrt{\frac{\delta t}{T+\delta t}} \sum_{t=0}^{T} |\psi^i(t) \rangle \otimes |t \rangle$, with $|\psi^i(t) \rangle$ being the wave functions generated by an SSE propagation for the same realization of jumps. This proves the desired result, that the ground states of the Stochastic Feynman Clock Hamiltonians are history states of the stochastic trajectories.


\end{document}